\input{aipcheck}

\documentclass[
    ,final            
  ]
  {aipproc}

\layoutstyle{6x9}

\def\ie{{\em i.e.,} }

\begin{document}

\title{Effect of plasma composition on accretion on to black holes}


\author{Indranil Chattopadhyay}{
  address={ARIES, Manora Peak, Nainital-263129, India}
}


%

\begin{abstract}
Matter makes a transition from non-relativistic to relativistic regime, as it falls onto a black hole. 
We employ a relativistic equation of state, abbreviated as RC, to study multi-species fluid flow around
black holes. We show that pair-plasma fluid around a black hole is thermally not relativistic. In order to make
it relativistic, a finite baryon loading is necessary. As a consequence of this, pair-plasma flow do not suffer
centrifugal pressure driven shock in accretion. However, fluid with finite baryon content may undergo shock transition.
\end{abstract}

\maketitle
\section{Assumptions and equations of motion}
Studies on accretion on to black holes are pursued in order to explain the radiation emitted, or formation of jets from AGNs or X-ray binaries. 
The inner boundary condition of black hole accretion, allows only sub-Keplerian matter to cross the horizon with the speed of light $c$.
Thus, accretion onto a black hole is necessarily
trans-relativistic and transonic. A fluid is relativistic, if its bulk velocity $v{\sim}c$, and/or, its thermal energy is
comparable to its rest energy.
A thermally relativistic fluid is indicated by its adiabatic index $\Gamma{\sim}4/3$, and a non-relativistic fluid by 
$\Gamma \sim 5/3$.
The fixed $\Gamma$-law
(hereafter ID) equation of state (EoS) is inadequate and ignores the temperature dependence of $\Gamma$ [1, 2, 3]
Instead a relativistic EoS [4, 5, 6, 3] 
should be used to describe such a fluid.
Solutions of relativistic fluid around black holes
have been undertaken by a number authors [7, 8, 9, 10, 11, 12, 13].
Apart from few authors [8, 9, 12], 
majority of them used the ID EoS.
In this paper, sonic point properties and solutions of
multi species, rotating fluid has been studied. We show that, pair plasma fluid is the least relativistic compared to
fluids with finite baryonic content.

To simplify the problem let us consider, an inviscid, rotating, non-magnetic fluid accreting around a Schwarzschild
black hole. The fluid is assumed to be composed of electrons, positrons and protons, and the flow geometry is wedge
shaped. The equations of motions in steady state are,

\begin{equation}
u^r\frac{du^r}{dr}+\frac{1}{r^2}-(r-3)u^{\phi}u^{\phi}=-\left(1-\frac{2}{r}+u^ru^r\right)\frac{1}{e+p}
\frac{dp}{dr}, \label{mom}
\end{equation}
\begin{equation}
\frac{de}{dr}-\frac{e+p}{n}\frac{dn}{dr}=0, \label{entro}
\end{equation}
\begin{equation}
\frac{1}{n}\frac{dn}{dr}=-\frac{2}{r}-\frac{1}{u^r}\frac{du^r}{dr}. \label{cont}
\end{equation}
Here, $u^{\mu}$s are the components of 4-velocity. And $n$, $p$, $e$ are the total particle density, the pressure,
and the energy density, respectively.
The system of units used is $G=M=c=1$.
The most commonly used equation of state (EoS) is ID $e=\rho c^2+p/(\Gamma -1)$.
However, we employ a relativistic EoS which is abbreviated as RC [3], 
\begin{equation}
e=\rho c^2 +p\left(\frac{9p+3\rho c^2}{3p+2\rho c^2}\right), \label{rceos}
\end{equation}
The total particle number density is $n=n_e+n_++n_p$, where, $n_e$, $n_+$ and $n_p$ are
the electron, the positron and the proton number densities, respectively.
Moreover, mass density $\rho=\Sigma n_im_i=n_em_e\left\{ 2-\xi \left(1-1/\eta\right)\right\}$, where $\xi=n_p/n_e$
and $\eta=m_e/m_p$. Eqs. (1-3) along
with Eq.\eqref{rceos} are simplified and we get two dependent equations of $dv/dr$ and $d{\Theta}/dr$,
where $v^2=\left(-u_ru^r/u_tu^t\right)/\left(1-v^2_{\phi}\right)$,
$v^2_{\phi}=-u_{\phi}u^{\phi}/u_tu^t$, and $\Theta=kT/(m_ec^2)$. Moreover, the two constants of motion,
specific energy and angular momentum of the flow are,
\begin{equation}
{\cal E}=\frac{(f+2{\Theta})u_t}{(2-\xi+\xi/\eta)}, ~~ {\lambda}=-\frac{u_{\phi}}{u_t}, \label{cons}
\end{equation}
where $f=(2-\xi)\left[1+\Theta\left(\frac{9\Theta+3}{3\Theta+2}\right)\right]
+\xi\left[1/\eta+\Theta\left(\frac{9\Theta+3/\eta}{3\Theta+2/\eta}\right)\right]$. 

\begin{figure}
\hskip -2.0cm
  \includegraphics[height=.24\textheight]{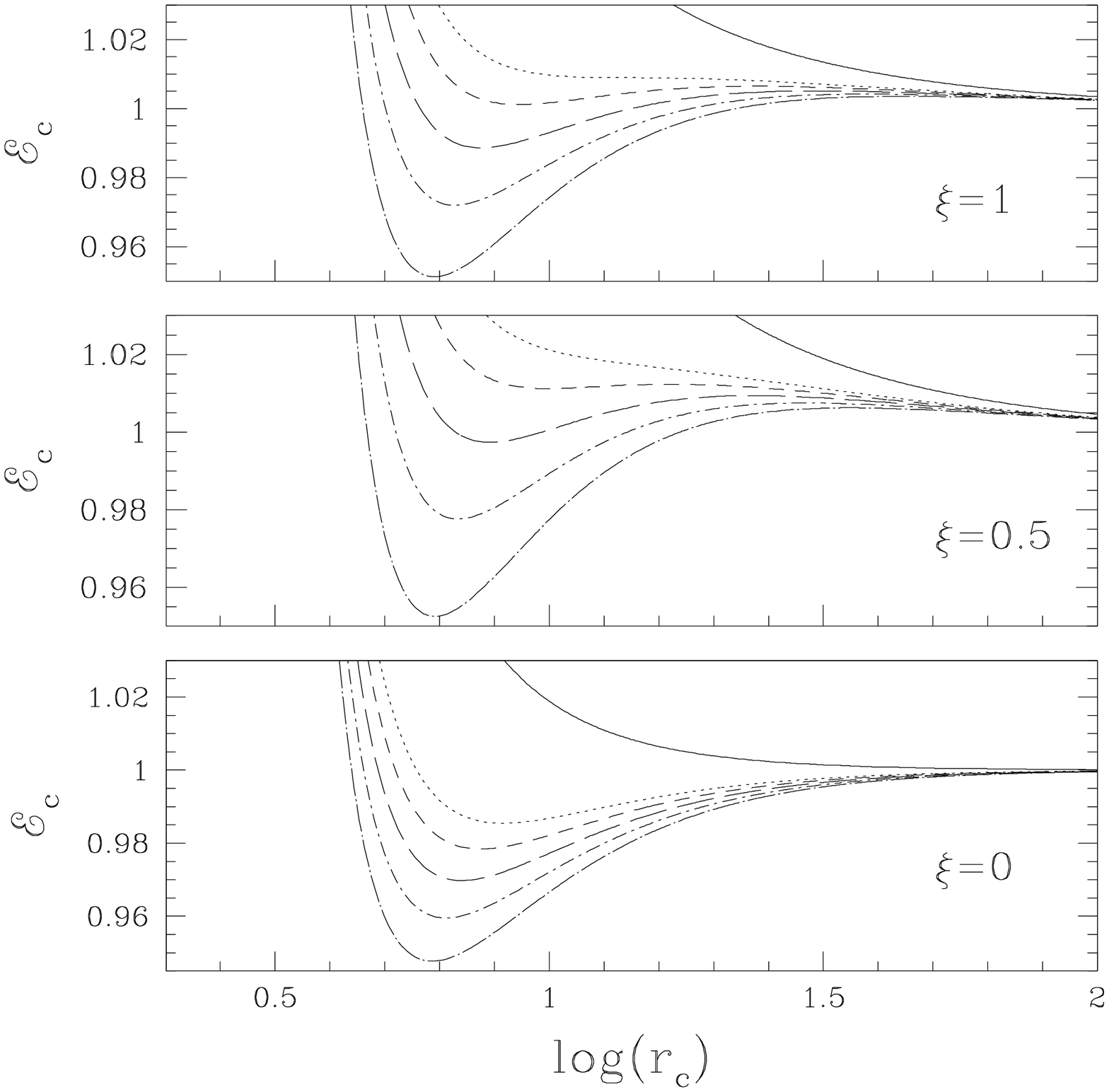}

\hskip -2.0cm (a)

\hskip 2.0cm
 \includegraphics[height=.24\textheight]{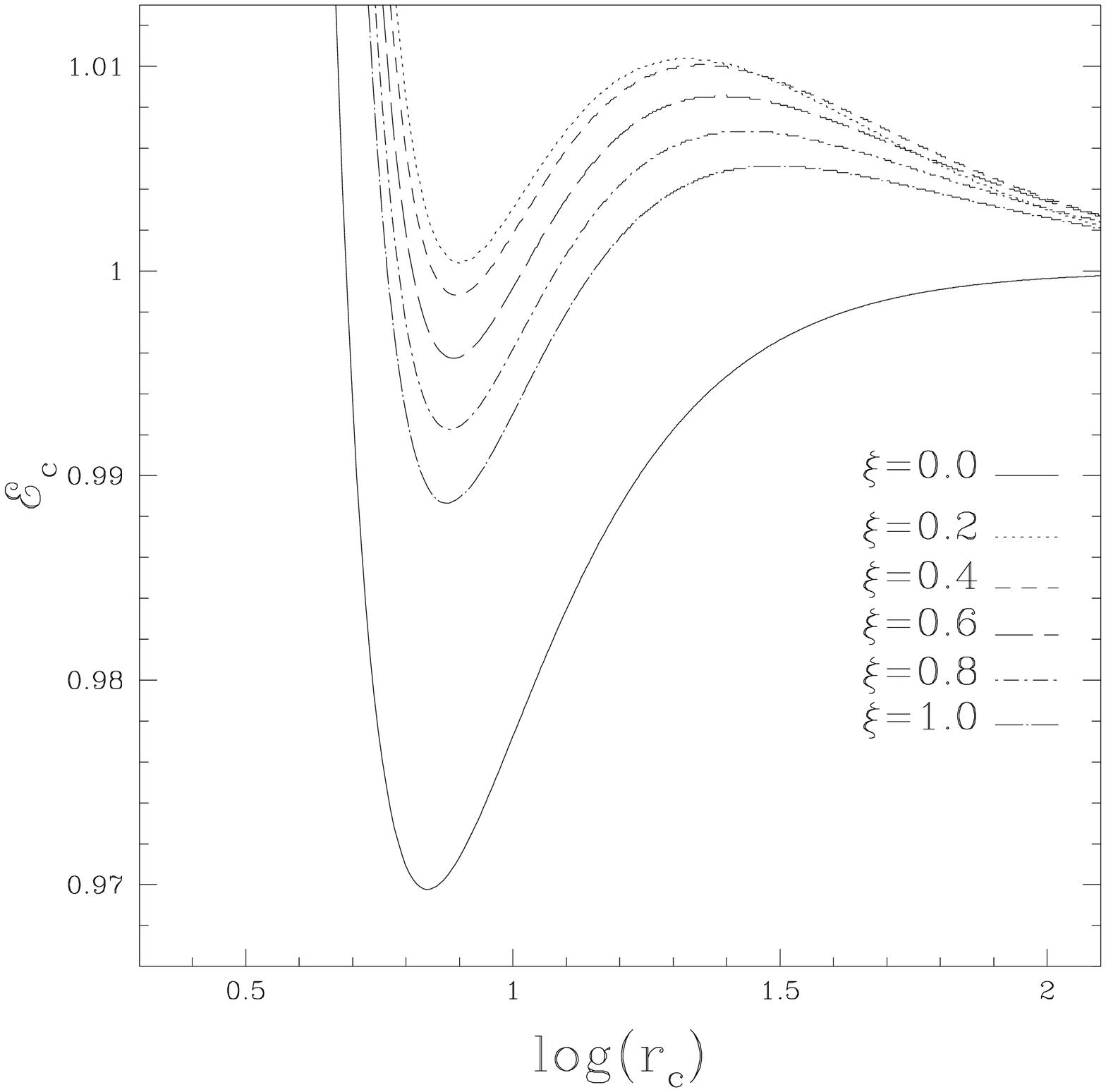}

\hskip -2.0cm (b)

\caption{${\cal E}_c$ is plotted with $log(r_c)$. (a) Each curve
parametrized by $\lambda=$ $0.0$ (solid), $2.8$ (dotted), $3.0$ (dashed), $3.2$ (long-dashed),
$3.4$ (dashed-dotted) and $3.6$ (long dashed-dotted), for three type of fluid given by
$\xi=1.0$ (top), $0.5$ (middle) and $0.0$ (bottom). (b) Each curve represents fluid of different composition,
but for same $\lambda=3.2$.}
\end{figure}

\begin{figure}
\hskip -2.0cm
  \includegraphics[height=.23\textheight]{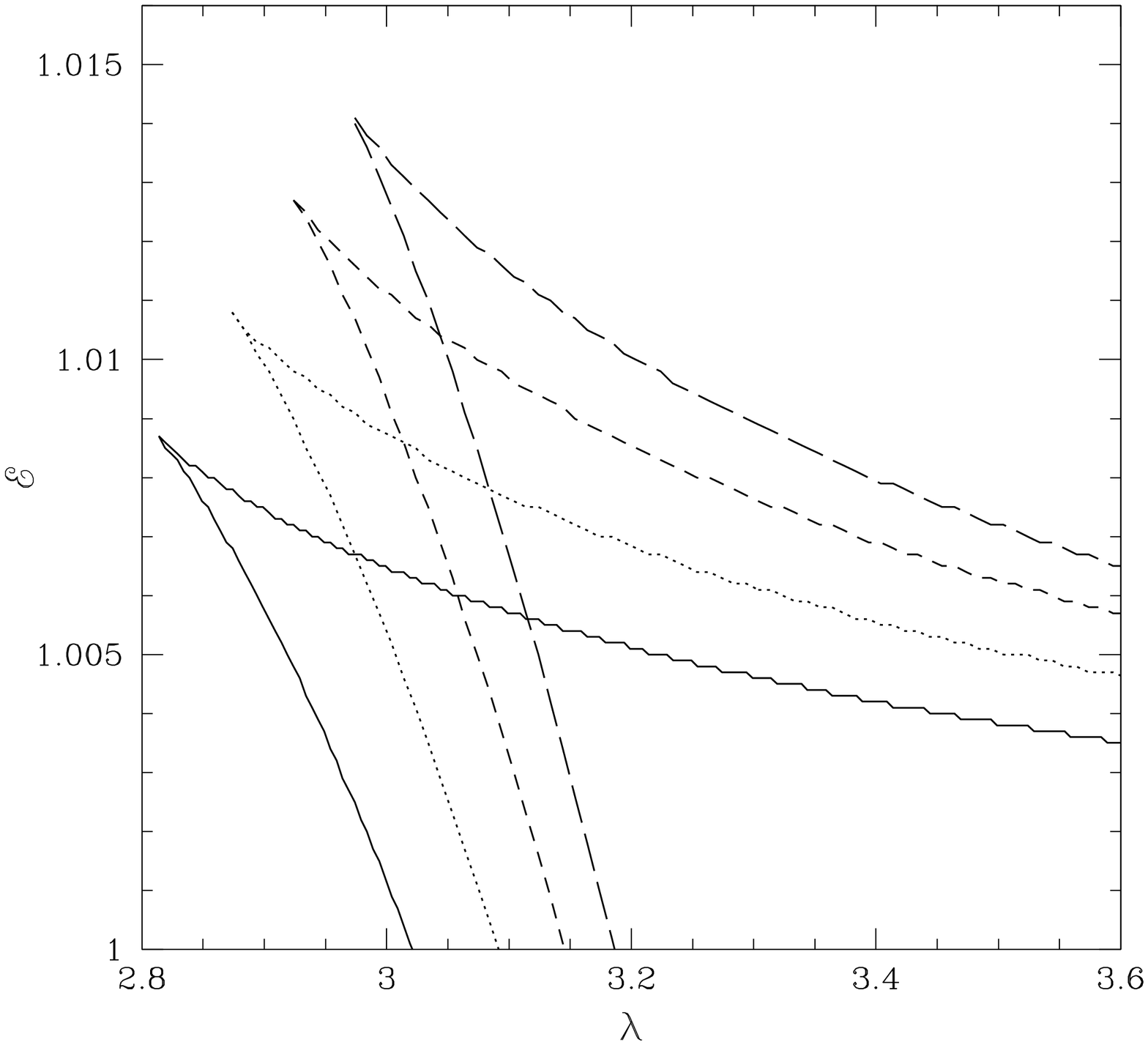}

\hskip -2.0cm (a)

\hskip 2.0cm
 \includegraphics[height=.23\textheight]{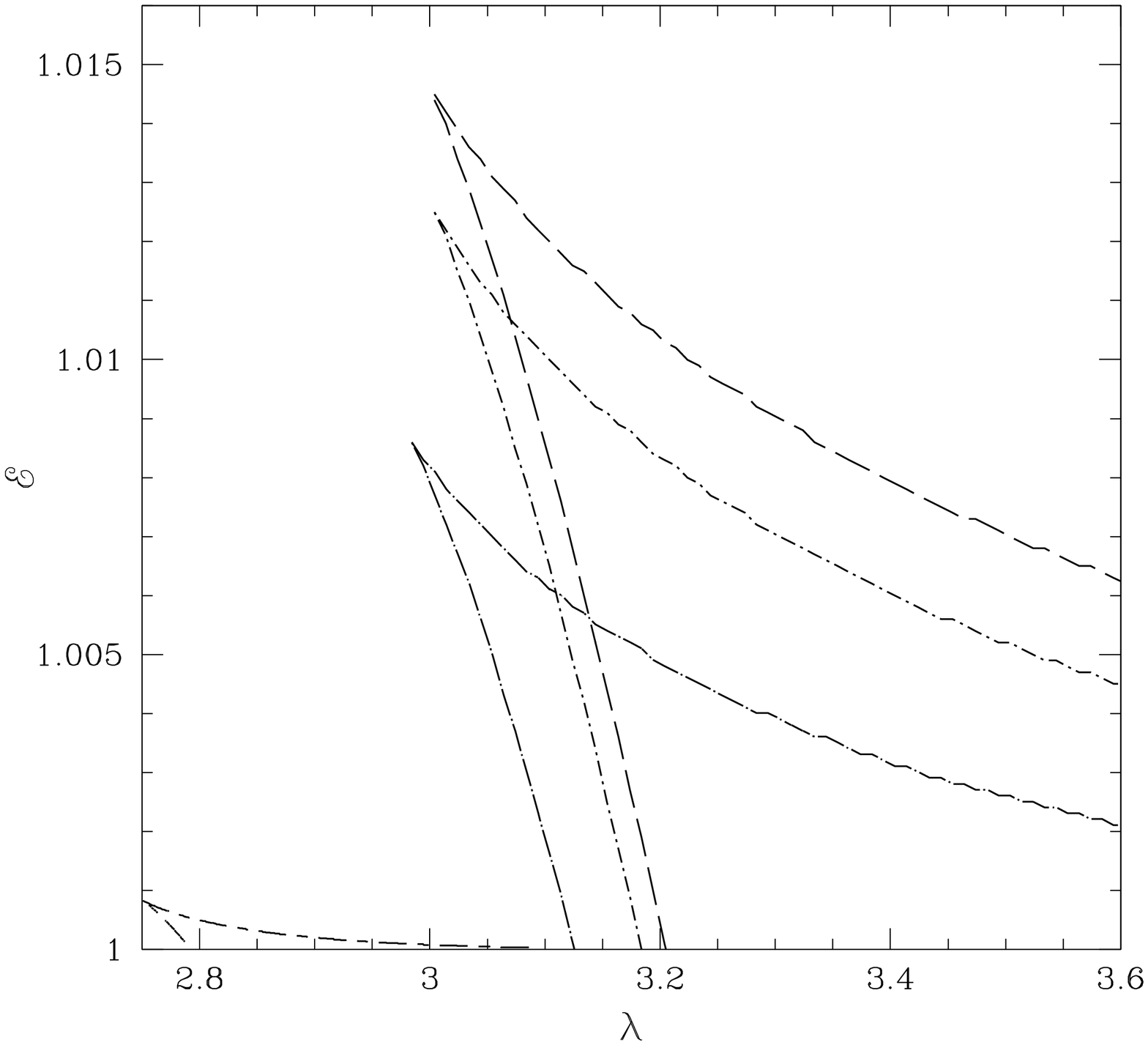}

\hskip -2.0cm (b)

\caption{Parameter space (${\cal E}$,$\lambda$)for three sonic point. Each curve
represents, (a) $\xi=1.0$ (solid), $0.8$ (dotted), $0.6$ (dashed), $0.4$ (long-dashed). (b)
$\xi=0.2$ (long dashed), $0.1$ (dashed dotted), $0.05$ (long dashed-dotted), $0.01$ (long-short dash).}
\end{figure}

\begin{figure}
\hskip -0.6cm
  \includegraphics[height=.2\textheight]{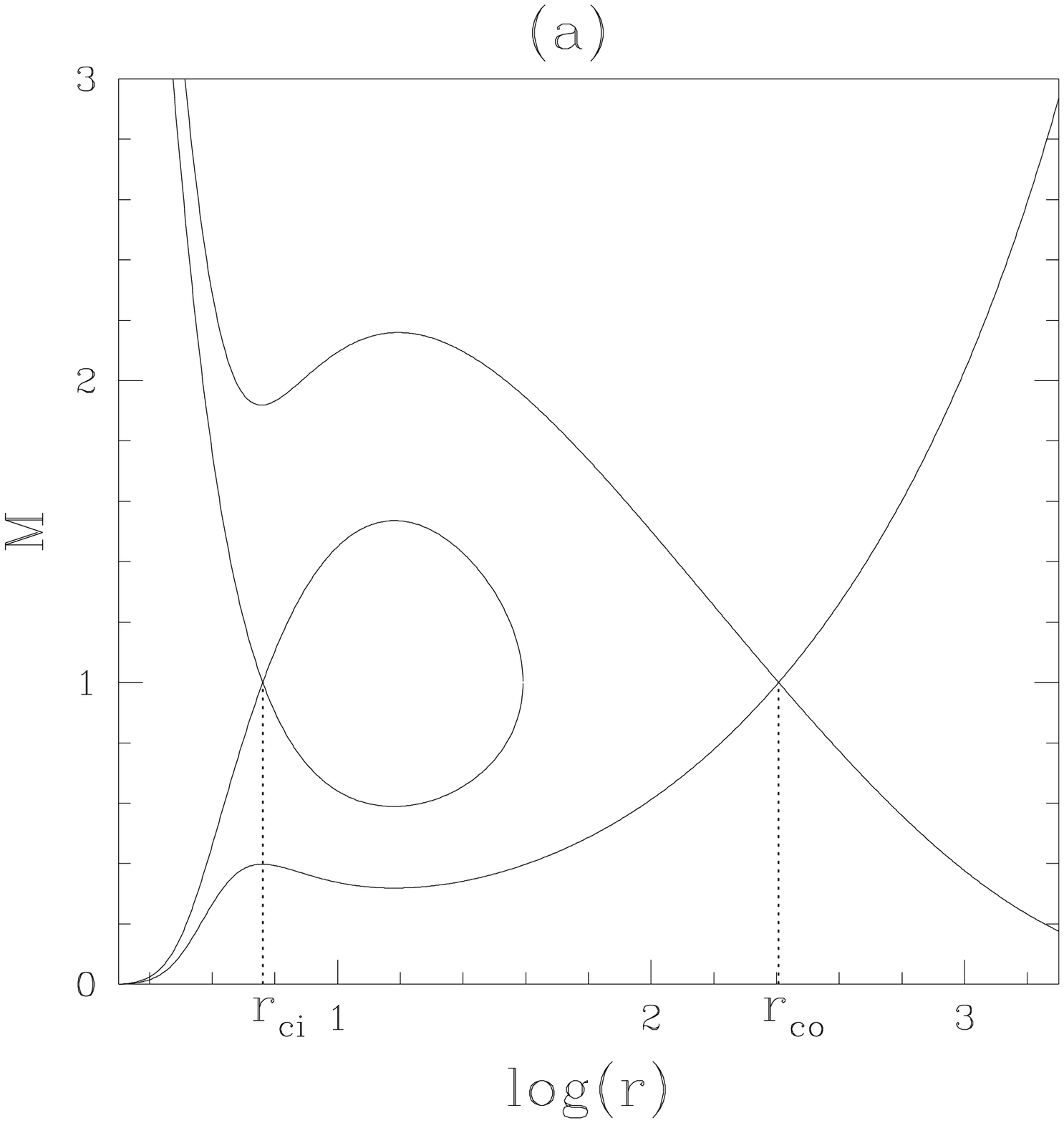}

%
\hskip 1.2cm
 \includegraphics[height=.2\textheight]{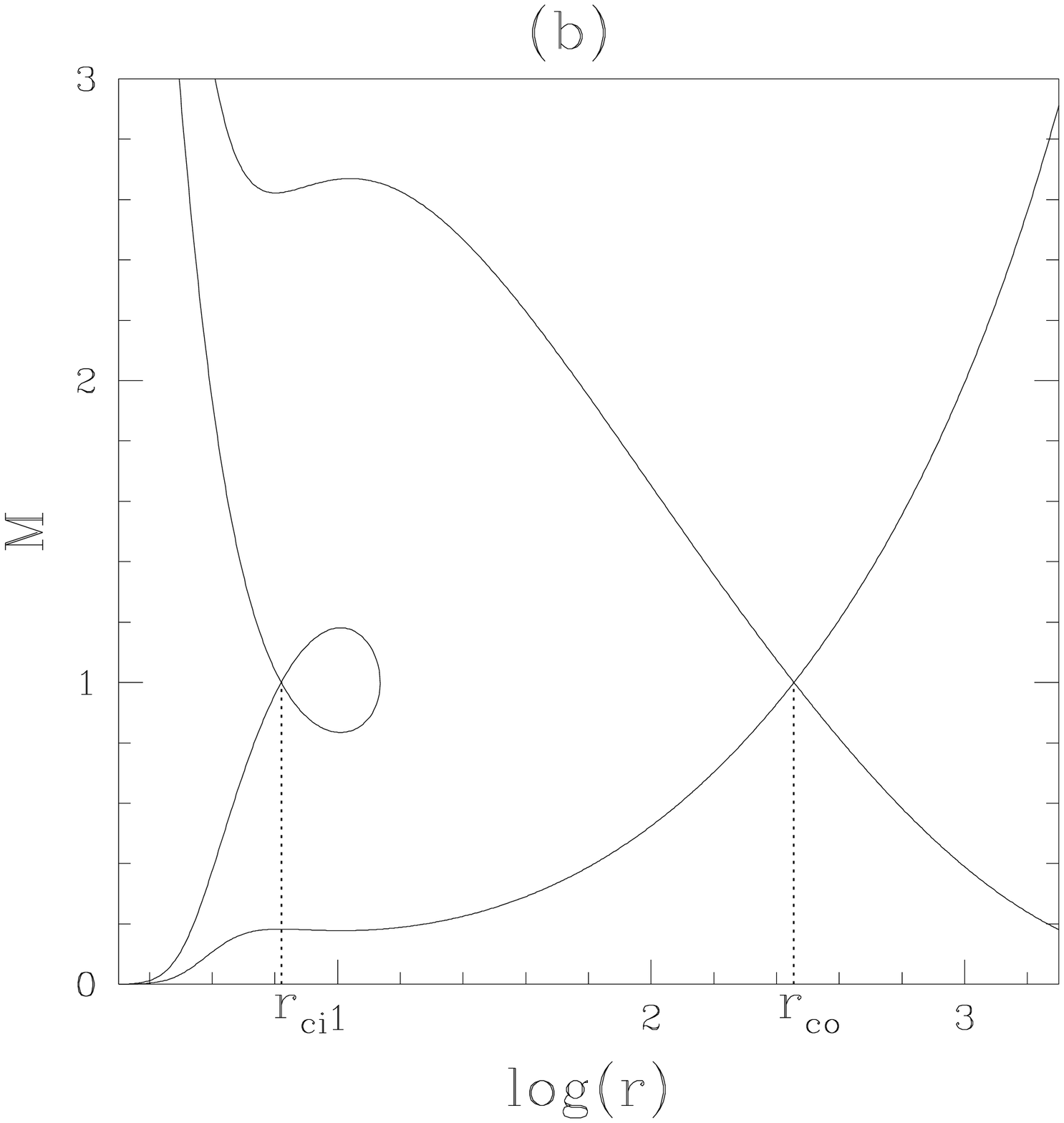}

%
\hskip 1.0cm
 \includegraphics[height=.2\textheight]{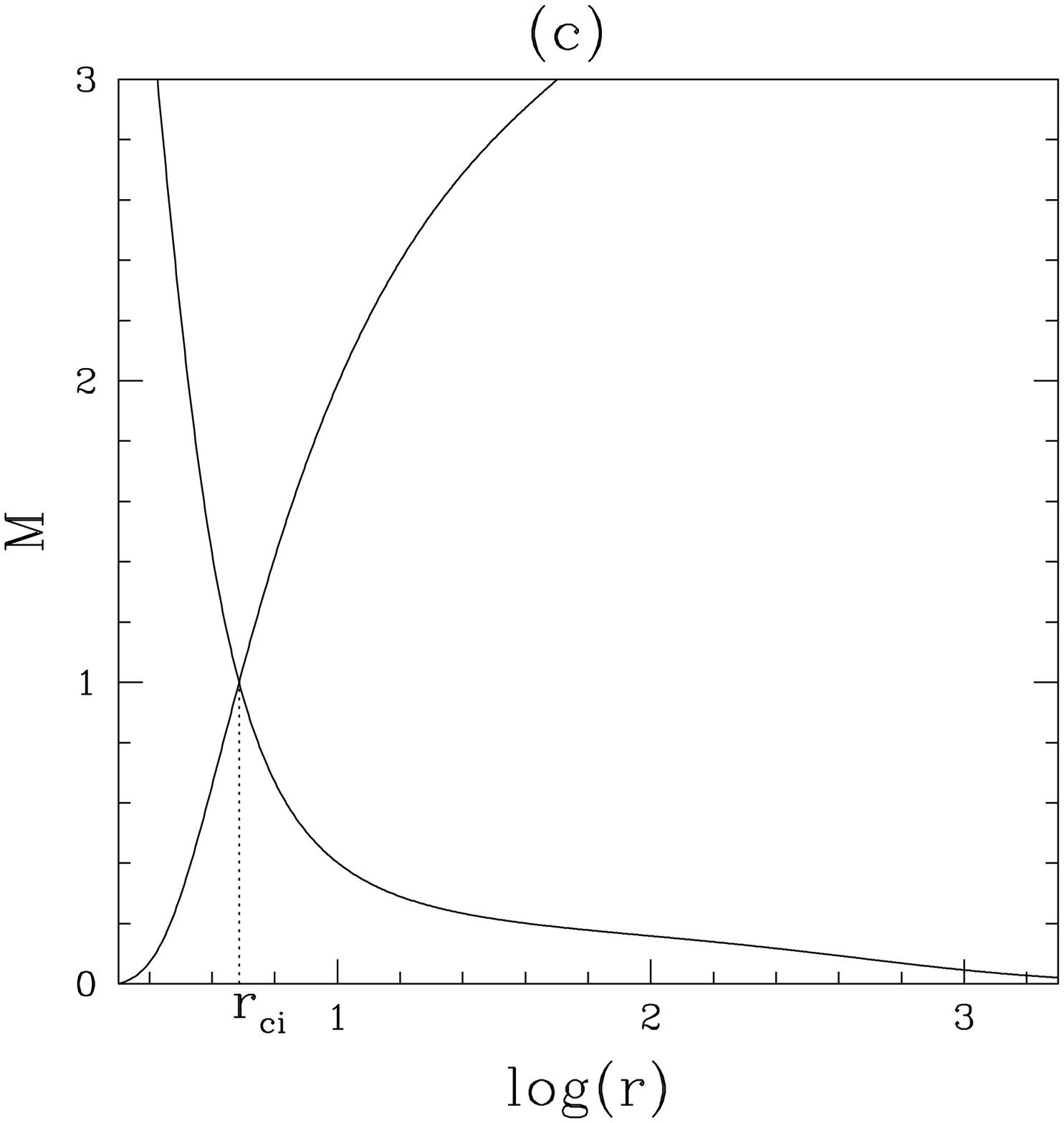}

\caption{Mach number $M$ is plotted with $log(r)$ for fluids of (a) $\xi=1.0$, (b) $\xi=0.5$ and (c) $\xi=0.0$, and for
same
${\cal E}=1.00096$ and $\lambda=3.2$. The position of inner $r_{ci}$, and outer $r_{co}$ X-type
sonic points are shown in the figure.}
\end{figure}
\section{Results and summary}

Since accretion onto black holes is necessarily transonic, therefore the fluid will make a transition from subsonic to
supersonic velocity at the sonic point $r_c$. The specific energy ${\cal E}_c$ at
$r_c$ is presented in Fig. 1, which
shows that sonic point properties depend on both $\lambda$ and $\xi$. While fluid with finite baryons
admit 3 sonic points in significant portion of the parameter space (top and middle panel of Fig. 1a),
but purely pair plasma (bottom panel of Fig. 1a) do not show the existence of 3 sonic point. This fact is also
vindicated by Fig. 1b.
The parameter space ${\cal E}-\lambda$, for 3 sonic point is constructed for each $\xi$ by
connecting the extrema of ${\cal E}_c$---$r_c$ plot, and is represented in Fig. 2. 
The  proton content decreases as $\xi$ decreases,
and the fluid becomes more relativistic and the flow becomes more energetic, however,
as $\xi{\rightarrow}0$, the fluid becomes too cold, and the parameter space for three sonic point shrinks to zero.
In Fig. (3a-c), the solution topologies are plotted for (a) $\xi=1.0$, (b) $\xi=0.5$ and (c) $\xi=0.0$,
where (${\cal E},\lambda$)=($1.00096,3.2$). It clearly shows that, the actual solution strongly depends
on the composition of the fluid even for same ${\cal E}$ and $\lambda$, and vindicates our
parametric study that pair plasma ($\xi=0.0$) has no multiple sonic point.
If multiple X-type sonic points exist, then
the possibility of shock wave enhances. We check for standing adiabatic shock conditions for relativistic fluid [1].
In Fig. 4a, we plot the shock location 
$r_s$ with ${\cal E}$ for $\lambda=3.3,~3.4$, and for different composition of the fluid, \ie $\xi=1.0$ (solid) and
$0.8$ (dotted).
In Fig. 4b, the parameter space for standing shock is plotted for different $\xi$s. The shock parameter space moves up
and rightward for $1{\geq}\xi{\geq}0.2$, however, moves down and leftward for
$\xi<0.2$. In Fig. 4c, the full parameter space for solution of a fluid with ID EoS, and an
electron-proton fluid with RC EoS is compared, which signifies the importance of using correct physics.

\begin{figure}
\hskip -1.0cm
  \includegraphics[height=.23\textheight]{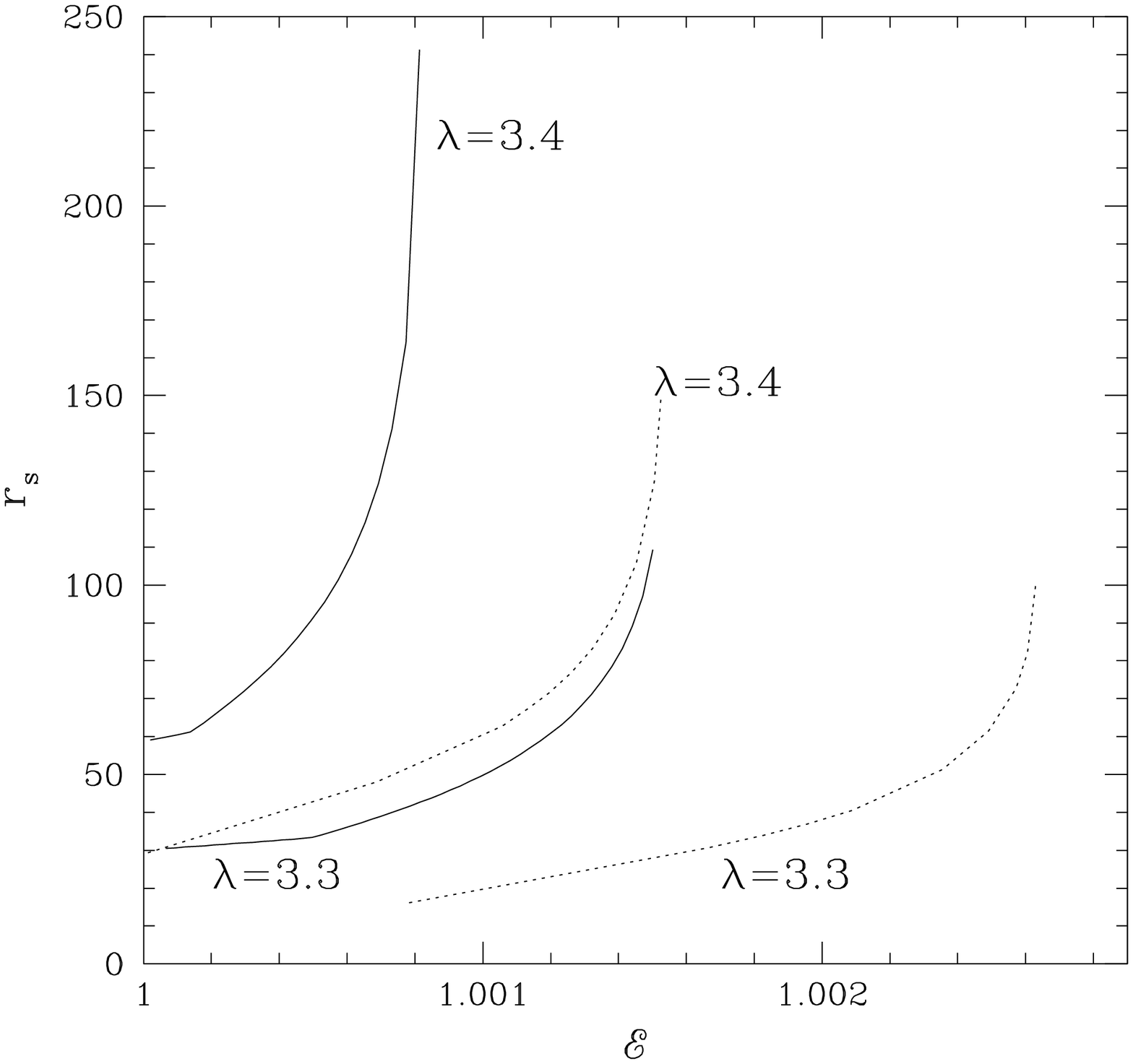}

\hskip -2.0cm (a)

\hskip 2.0cm
 \includegraphics[height=.23\textheight]{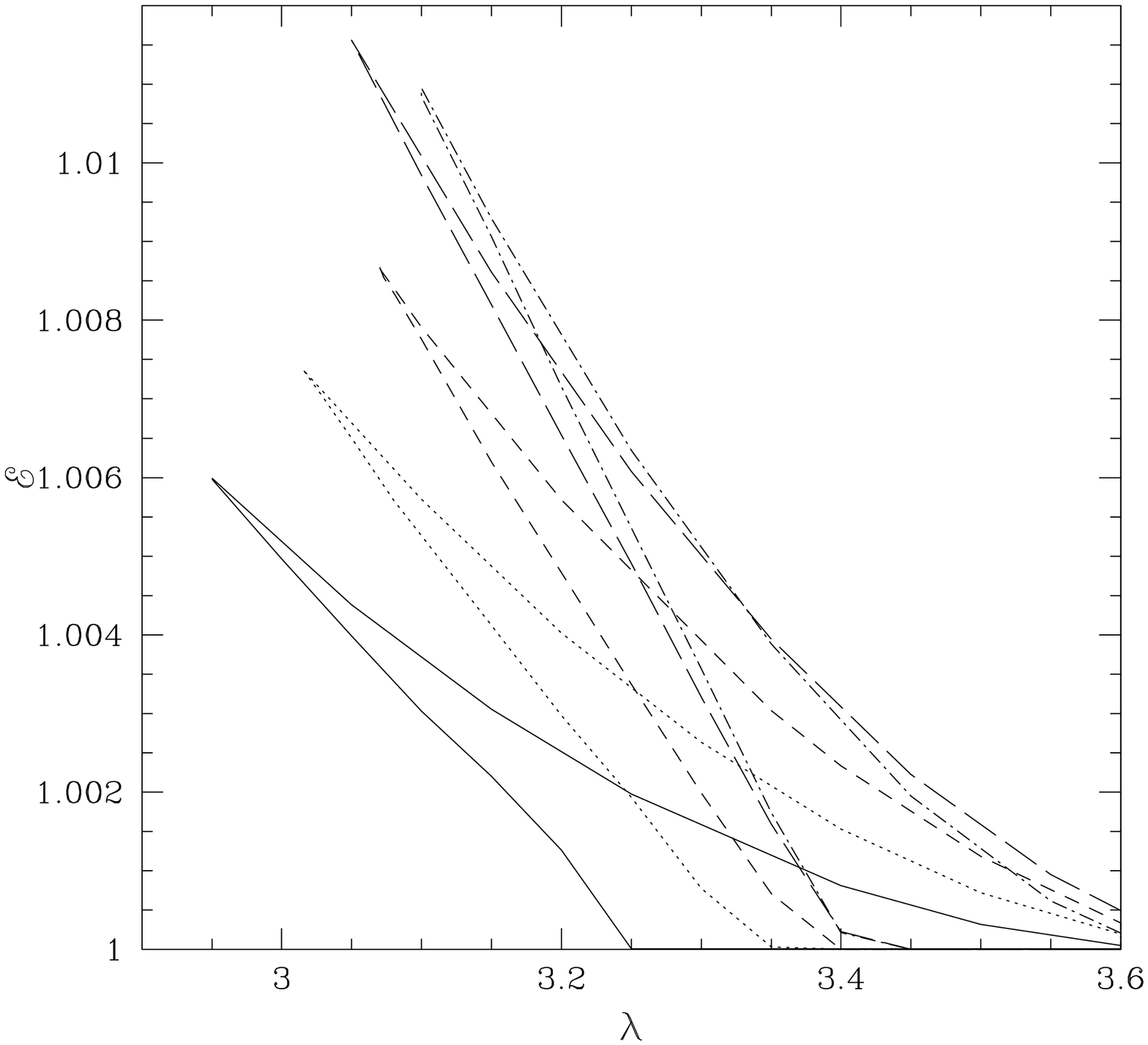}

\hskip -2.0cm (b)

\hskip 1.5cm
  \includegraphics[height=.23\textheight]{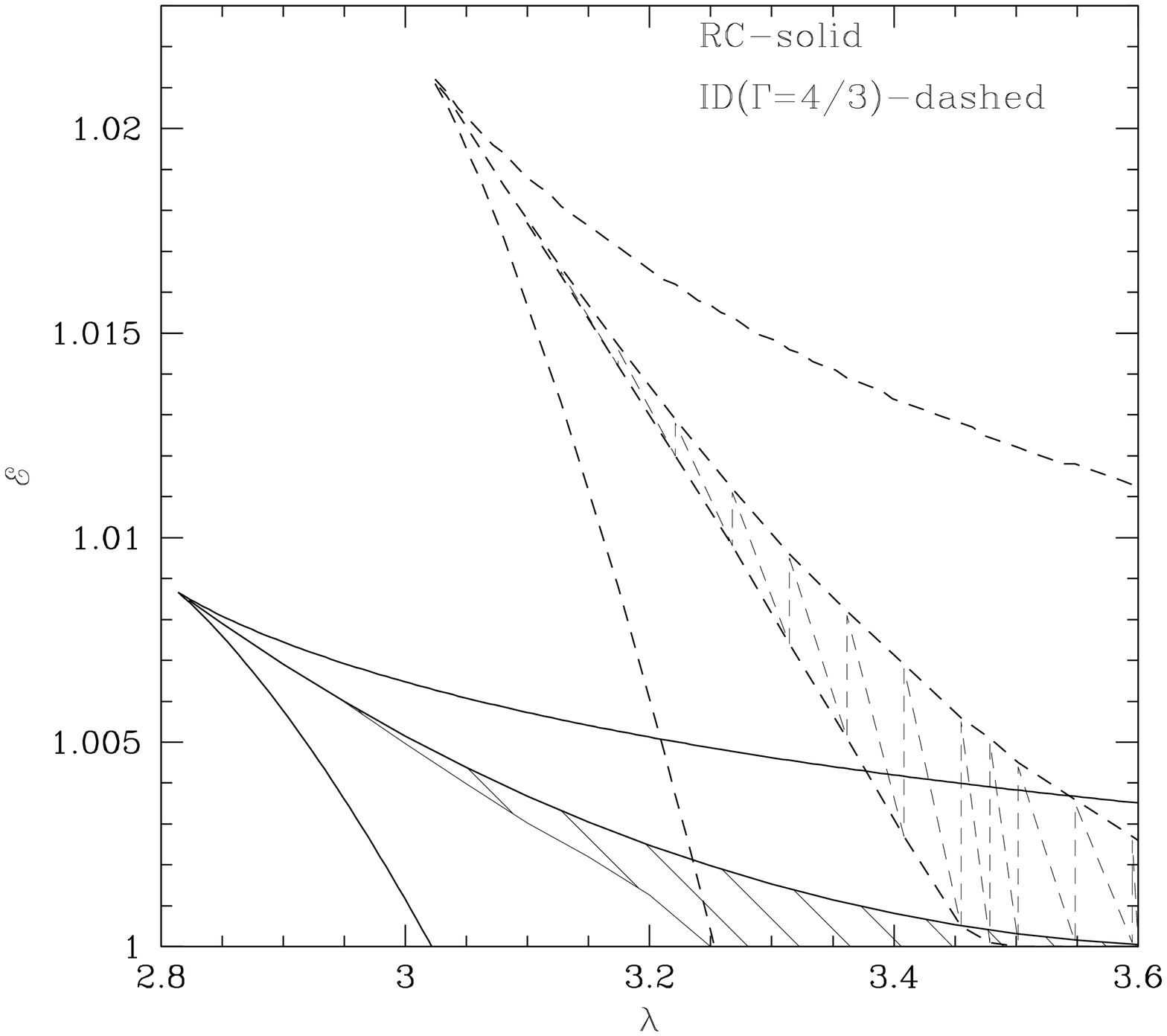}

 \hskip -1.0cm (c)

  \caption{(a) Variation of shock location $r_s$ with ${\cal E}$ for two values of $\lambda$s
($3.3,3.4$), and for different values of $\xi=1.0$ (solid) and $\xi=0.8$ (dotted). (b) Parameter
space (${\cal E},{\lambda}$) which admits shock solution, $\xi=1$ (solid), $0.8$ (dotted), $0.6$
(dashed), $0.4$ (long dashed), $0.2$ (dashed-dotted). (c) The bounded domain allows 3 sonic points,
the shaded one is for shocked fluid, for electron-proton fluid (solid) and with ID EoS (dashed).}
\end{figure}

It is clear that, fluid with same energy and angular momentum but of different composition, not only produce different
solution topologies,
even their observable properties such as shock also depend strongly on the composition.
And if indeed, these shocks
explain the high energy spectra of black hole candidates [14, 15], 
or generation of jets [16, 17],
then it is of utmost importance that we use correct physics. Since most of the hard radiation
originates closer to the horizon, hence modelling the inner region with relativistic
EoS as well as consideration of correct fluid composition is very important to understand the physics
of the inner region of an accretion disc.

\vskip 0.5cm
\hskip 5.0cm  {\bf REFERENCES} \\
\noindent 1.  A. H. Taub, \emph{Phys. Rev.}, \textbf{74}, 328 (1948). \\
\noindent 2.  A.~Mignone, T.~Plewa, and G.~Bodo, \emph{ApJS}, \textbf{160}, 199 (2005). \\
\noindent 3.  D.~Ryu, I.~Chattopadhyay, and E.~Choi, \emph{ApJS}, \textbf{166}, 410 (2006). \\
\noindent 4.  S.~Chandrasekhar, in \emph{An Introduction to the Study of Stellar Structure}, Dover, NewYork, (1938). \\
\noindent 5.  J.~L. Synge, in \emph{The Relativistic Gas} North Holland, Amsterdam (1957). \\
\noindent 6.  W.~G. Mathews, \emph{ApJ}, \textbf{165}, 147 (1971). \\
\noindent 7.  F.~C. Michel, \emph{Ap\&SS}, \textbf{15}, 153 (1972). \\
\noindent 8.  G.~R. Blumenthal, and W.~G. Mathews, \emph{ApJ}, \textbf{203}, 714 (1976). \\
\noindent 9.  J.~Fukue, \emph{PASJ}, \textbf{39}, 309 (1987). \\
\noindent 10. S.~K. Chakrabarti, \emph{MNRAS}, \textbf{283}, 325 (1996). \\
\noindent 11. T.~K. Das, \emph{A\&A}, \textbf{376}, 697 (2001). \\
\noindent 12. Z.~Meliani, C.~Sauty, K.~Tsinganos, and  N.~Vlahakis, \emph{A\&A}, \textbf{425}, 773 (2004). \\
\noindent 13. K.~Fukumura and D.~Kazanas, \emph{ApJ}, \textbf{669}, 85 (2007). \\
\noindent 14. S.~K. Chakrabarti and  L.~Titarchuk, \emph{ApJ}, \textbf{455}, 623 (1995). \\
\noindent 15. S.~K. Chakrabarti and S.~Mandal, \emph{ApJ}, \textbf{642}, L49 (2006). \\
\noindent 16. I.~Chattopadhyay and S.~Das, \emph{New A.}, \textbf{12}, 454 (2007). \\
\noindent 17. S.~Das and I.~Chattopadhyay, \emph{New A.}, \textbf{13}, 549 (2008).

\end{document}